\documentclass[twocolumn]{jpsj3}
\usepackage{txfonts}
\usepackage{graphicx}
\usepackage{dcolumn}
\usepackage{bm}
\usepackage[usenames,dvipsnames]{xcolor}
\usepackage{ulem}
\usepackage{braket}
\usepackage{amsmath}

\title{Magnetic Vortex induced by Nonmagnetic Impurity in Ferromagnets:\\ 
Magnetic Multipole and Toroidal around the Vacancy}

\author{Satoru Hayami$^1$, Hiroaki Kusunose$^2$, and Yukitoshi Motome$^3$}
\inst{
 $^1$Faculty of Science, Hokkaido University, Sapporo 060-0810, Japan \\
 $^2$Department of Physics, Meiji University, Kawasaki 214-8571, Japan \\
 $^3$Department of Applied Physics, University of Tokyo, Tokyo 113-8656, Japan
 }

\abst{
We report our theoretical study on nucleation of a magnetic vortex around a nonmagnetic impurity doped into ferromagnets. 
The mechanism lies in the asymmetric Dzyaloshinskii-Moriya interaction arising from the breaking of spatial inversion symmetry by the impurity. 
By using the spin-wave analysis and Monte Carlo simulations, we show that the asymmetric interaction induces a magnetic vortex with nonzero vorticity $l=+1$. 
The vortex is stabilized even for less frustration in exchange interactions and in the absence of an external magnetic field. 
We also find that the magnetic vortex is characterized by magnetic multipoles according to its vorticity and helicity. 
We demonstrate a potential realization of such a magnetic vortex by considering a monolayer ferromagnet on a nonmagnetic substrate, which results in the magnetic monopole and toroidal dipole. 
}

\begin{document}
\maketitle

Topological defects in magnets have attracted much interest, as they give rise to nontrivial quantum states and dynamics~\cite{Anderson_PhysRevLett.38.508,kosevich1990magnetic,bar1994dynamics,Meier_PhysRevX.7.041014}. 
For example, magnetic Skyrmions, which are particle-like topological solitons~\cite{skyrme1962unified}, have long been studied in various fields of condensed matter physics, such as liquid $^3$He-A~\cite{Anderson_PhysRevLett.38.508} and atomic Bose-Einstein condensates~\cite{Ruostekoski_PhysRevLett.86.3934,Leslie_PhysRevLett.103.250401}. 
Since the discovery in B20 compounds~\cite{Muhlbauer_2009skyrmion,yu2010real}, the topologically protected spin textures have been extensively studied in chiral magnets~\cite{nagaosa2013topological}. 
In particular, their control is an important issue, since they are potentially used as information carriers~\cite{romming2013writing,fert2013skyrmions,fert2017magnetic}. 

Spin vacancies by replacing magnetic ions with nonmagnetic ones provide a way to control such topological spin textures. 
For example, nonmagnetic impurities doped into triangular-lattice antiferromagnets under an external magnetic field induce a noncoplanar magnetic structure with nonzero spin scalar chirality, due to an effective positive biquadratic interaction between magnetic ions activated by the impurity doping~\cite{Wollny_PhysRevLett.107.137204,Sen_PhysRevB.86.205134,maryasin2015collective,Maryasin_PhysRevLett.111.247201}.
Another example is found in frustrated magnets near a Lifshitz transition between incommensurate spiral ordering and the ferromagnetic ordering; 
a nonmagnetic impurity nucleates a magnetic vortex in a finite range of magnetic field above the bulk saturation field~\cite{Lin_PhysRevLett.116.187202,Hayami_PhysRevB.94.174420,batista2016frustration}.

Motivated by these studies, in this Letter, we propose another intriguing mechanism for nucleation of a magnetic vortex by introducing a nonmagnetic impurity. 
We focus on the lowering of structural symmetry by the vacancy. 
Considering the ferromagnetic state on a triangular lattice, we show that the symmetry reduction turns on the Dzyaloshinskii-Moriya (DM) interaction~\cite{dzyaloshinsky1958thermodynamic,moriya1960anisotropic} around the nonmagnetic impurity, which induces a magnetic vortex with nonzero vorticity. 
By the spin-wave analysis and Monte Carlo simulations, we find that the magnetic vortex is stabilized even for less frustration in exchange interactions at zero magnetic field. 
Furthermore, considering the monolayer triangular ferromagnet on a nonmagnetic substrate, we classify the impurity-induced magnetic vortices in terms of the cluster magnetic multipoles~\cite{EdererPhysRevB.76.214404,Spaldin_0953-8984-20-43-434203,Yanase_JPSJ.83.014703,Hayami_PhysRevB.90.024432,hayami2016emergent,Gao_PhysRevB.97.134423}. 
We find that the magnetic toroidal dipole becomes dominant for large positive anisotropic and small Rashba-type exchange couplings, while the magnetic monopole is favored in the rest parameter region. 

Let us consider a spin model on a triangular lattice. 
The following analysis can be straightforwardly extended to other two-dimensional lattices, such as a square lattice. 
The Hamiltonian is given by
\begin{align}
\label{eq:Ham}
\mathcal{H}= -\sum_{i,j} J_{ij} \bm{S}_i \cdot \bm{S}_j -A \sum_i (S_i^z)^2, 
\end{align}
where $\bm{S}_i=(S_i^x,S_i^y,S_i^z)$ represents a classical localized spin with $|\bm{S}_i|=1$. 
The first term describes the Heisenberg-type exchange interactions. 
We here consider the ferromagnetic exchange interaction for nearest neighbors, $J_1>0$, and the antiferromagnetic one for third neighbors, $J_3<0$; other $J_{ij}$ are all taken to be zero.
We set $J_1=1$ as the energy unit and the lattice constant $a=1$ as the length unit. 
In the following, we focus on the parameter region of $|J_3|< J_1/4$ where the ground state is ferromagnetic. 
The second term represents the easy-axis spin anisotropy with $A>0$.

\begin{figure}[t!]
\begin{center}
\includegraphics[width=1.0 \hsize]{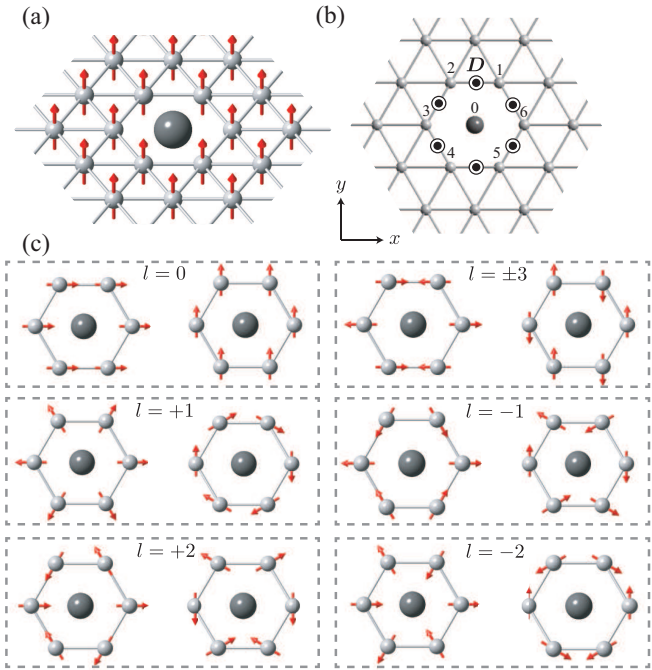} 
\caption{
\label{Fig:hole_ponti}
(Color online) 
(a) A nonmagnetic impurity in the triangular ferromagnet. 
(b) DM interactions between the spins surrounding the impurity in Eq.~(\ref{eq:HDM}). 
The DM vector is perpendicular to the plane. 
(c) Schematic in-plane magnetic structures classified by the vorticity $l$ around the impurity site. 
}
\end{center}
\end{figure}

For the model in Eq.~(\ref{eq:Ham}), we introduce a nonmagnetic impurity at the origin $\bm{r}_0=(0,0)$ represented by $\bm{S}_0=(0,0,0)$, as shown in Fig.~\ref{Fig:hole_ponti}(a). 
A similar situation was studied in Ref.~\citen{Lin_PhysRevLett.116.187202} under a magnetic field at $A=0$; 
it was shown that the nonmagnetic impurity can nucleate a magnetic vortex in the forced ferromagnetic state above the saturation field when $|J_3|>J_1/4$. 
This was ascribed to the spin canting by frustration between $J_1$ and $J_3$ that becomes conspicuous around the impurity. 
In this study, we discuss another mechanism for stabilizing a magnetic vortex that works even for less frustration $|J_3|<J_1/4$ and zero magnetic field. 
We focus on the breaking of spatial inversion symmetry by the impurity. 
It gives rise to additional asymmetric exchange interactions, the so-called DM interactions, in the presence of the relativistic spin-orbit coupling~\cite{dzyaloshinsky1958thermodynamic,moriya1960anisotropic}. 
Assuming that the spins neighboring to the nonmagnetic site are dominantly affected by the lowering of symmetry, the DM interactions are described by 
the following Hamiltonian: 
\begin{align}
\label{eq:HDM}
\mathcal{H}_{\rm DM} = 
-\sum_{ij} \bm{D}_{ij} \cdot (\bm{S}_i \times \bm{S}_j) = 
-\sum_{p} D (S_p^x S_{p+1}^y - S_p^y S_{p+1}^x), 
\end{align}
where the sum is taken for $\bm{r}_p= [\cos(p\pi /3 ),\sin(p\pi /3 )]$ ($p=1$-$6$) and $\bm{S}_{7} = \bm{S}_{1}$ [see Fig.~\ref{Fig:hole_ponti}(b)]. 
In Eq.~\eqref{eq:HDM}, the DM vector $\bm{D}$ is uniform on the six bonds and along the $z$ direction due to inversion and $C_6$ rotational symmetries around the impurity. 

As the total Hamiltonian given by Eqs.~(\ref{eq:Ham}) and (\ref{eq:HDM}) is invariant for a real-space sixfold rotation around the $z$ axis combined with an arbitrary spin rotation around the $z$ axis [$U(1)$ symmetry], the in-plane spin states at the six sites surrounding the impurity can be characterized by the vorticity $l$ ($0 \leq |l| \leq 3$) defined around the origin. 
The $(x, y)$-spin components of the magnetic structure with vorticity $l$ and helicity $\gamma$ are given as 
\begin{align} 
(S_p^x, S_p^y)\propto\left[\cos \left(\frac{\pi}{3}pl+ \gamma\right),\sin \left(\frac{\pi}{3}pl+ \gamma\right) \right], 
\label{eq:SxSy}
\end{align}
which are schematically shown in Fig.~\ref{Fig:hole_ponti}(c). 
The magnetic states with $|l|=1$ and 2 are accompanied with magnetic vortices. 
Note that the helicity $\gamma$ is arbitrary due to the presence of $U(1)$ symmetry.

\begin{figure}[t!]
\begin{center}
\includegraphics[width=1.0 \hsize]{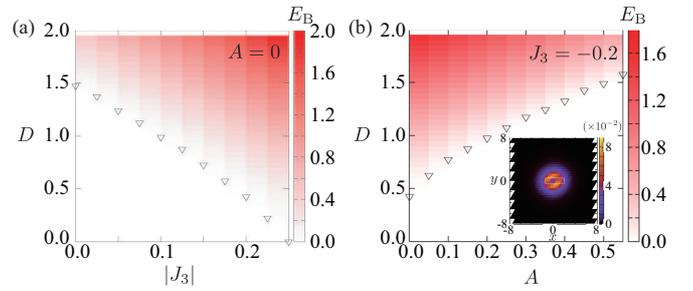} 
\caption{
\label{Fig:bindingenergy}
(Color online) (a) The contour of the binding energy $E_{\rm B}$ in the $|J_3|$-$D$ plane in the absence of $A$. 
The triangles represent the critical values of $D$. 
(b) The contour of $E_{\rm B}$ in the $A$-$D$ plane at $J_3=-0.2$. 
The inset shows the amplitude of the wave function for the vortex bound state with $l=+1$ for $D=0.8$ and $A=0.05$. 
}
\end{center}
\end{figure}

In order to clarify the magnetic instability around the impurity in the ferromagnetic background, we examine magnon excitations in the ferromagnetic state. 
For that purpose, we use the linear spin-wave theory by adopting the standard Holstein-Primakoff transformation, $S_i^z=S_i-a_i^{\dagger} a_i$, $S^- = \sqrt{2S_i} a^{\dagger}_i$, and $S^+= \sqrt{2S_i} a_i$, where $a_i$ is the boson operator at site $i$, and consider the low-density limit of magnons with $S_i = 1$.
The spin-wave Hamiltonians for Eqs.~(\ref{eq:Ham}) and (\ref{eq:HDM}) are given by  
\begin{align}
&\mathcal{H}=-\sum_{i,j}J_{ij} (a_i^{\dagger} a_j + a_j^{\dagger}a_i -a_i^{\dagger}a_i -a_j^{\dagger}a_j ) 
+2 A \sum_i a_i^{\dagger}a_i,
\\
&\mathcal{H}_{\rm DM}= {\rm i} D \sum_{p} (a_p^{\dagger}a_{p+1}-a_{p+1}^{\dagger}a_p), 
\end{align}
up to a constant.
By diagonalizing $\mathcal{H}+\mathcal{H}_{\rm DM}$, we can obtain a bound state around the impurity when the impurity doping lowers the energy. 
The binding energy $E_{\rm B}$ is defined by the energy gain from the zero energy. 

Let us show how the bound state appears in the presence of $D$ at $A=0$, assuming that the ferromagnetic state polarized along the $z$ direction is the ground state. 
Figure~\ref{Fig:bindingenergy}(a) shows the binding energy $E_{\rm B}$ on the $|J_3|$-$D$ plane at $A=0$. 
We take a sufficiently large system size with $N=120^2-1$ spins. 
The result shows that the bound state is obtained for $D \gtrsim 1.5$ even in the absence of $J_3$, and the critical value of $D$ decreases as increasing $|J_3|$.
This is because the antiferromagnetic interaction $J_3$ favors spin canting. 
When $J_3$ reaches at the Lifshitz point from the ferromagnetic state to the spiral state, i.e., $|J_3|=J_1/4$, the critical $D$ becomes zero, as discussed for the model with $D=0$~\cite{Lin_PhysRevLett.116.187202}. 
We find that the eigenfunction for the bound state has the vorticity $l=+1$ (the vorticity changes its sign for $D<0$), whose spin pattern is described by Eq.~\eqref{eq:SxSy}. 

Such a vortex state with $l=+1$ is also realized for the nonzero easy-axis anisotropy, which avoids zero-energy excitations in the absence of the nonmagnetic impurity. 
We show $A$ and $D$ dependences of the binding energy $E_{\rm B}$ at $J_3=-0.2$ in Fig.~\ref{Fig:bindingenergy}(b). 
The critical $D$ becomes larger for larger $A$ because the spins become rigid and hard to be canted. 
We plot the amplitude of the real-space wave function for the vortex bound state with $l=+1$ for $D=0.8$ and $A=0.05$ in the inset of Fig.~\ref{Fig:bindingenergy}(b), which is well localized around the impurity. 

The spin-wave analysis indicates that the magnetic vortex with $|l|=1$ appears around the impurity even for less or no frustration $|J_3|<J_1/4$ by taking into account the local DM interaction originating from the inversion symmetry breaking by the impurity. 
More interestingly, the vortex is stabilized even in the absence of the external magnetic field, in contrast to the previous study~\cite{Lin_PhysRevLett.116.187202}. 

\begin{figure}[t!]
\begin{center}
\includegraphics[width=1.0 \hsize]{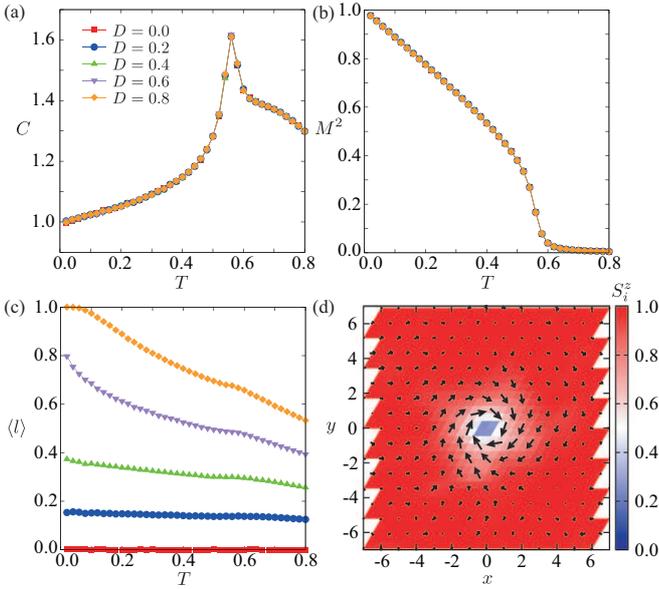} 
\caption{
\label{Fig:MCresult}
(Color online) $T$ dependences of (a) the specific heat $C$, (b) square of the magnetization, $M^2$, and (c) the vorticity $l$ around the impurity at $J_3=-0.2$ and $A=0.05$ for different $D$ obtained from the Monte Carlo simulations for the system size $N=60^2-1$. 
(d) Monte Carlo snapshot of the real-space spin configuration at $T=0.02$, $J_3=-0.2$, $D=0.8$, and $A=0.05$. 
The nonmagnetic impurity locates at the origin $x=y=0$, and a part of the whole system is shown. 
The arrows represent the $xy$ components of spins, and the color indicates the $z$ component.
}
\end{center}
\end{figure}

In order to examine the magnetic instability beyond the spin-wave analysis in the low-density limit of magnons, we perform Monte Carlo simulations. 
Our simulations are carried out with the standard Metropolis updates and the initial states are selected from random spin configurations. 
We also start the simulations from the fully-polarized ferromagnetic state at low temperature ($T$). 
We consider the triangular lattice with $N=60^2-1$ spins under the periodic boundary conditions, and perform $10^5-10^6$ Monte Carlo sweeps for measurements after $10^5-10^6$ steps for thermalization. 
The statistical errors are estimated from sixteen independent runs. 

We present the Monte Carlo results for several $D$ at $J_3=-0.2$ and $A=0.05$. 
Figure~\ref{Fig:MCresult}(a) shows $T$ dependence of the specific heat $C=(\braket{E^2}-\braket{E}^2)/T$, where $\langle \cdots \rangle$ is the thermal average, $E$ is the internal energy, and the Boltzmann constant is set to be unity. 
The peak of $C$ at $T \sim 0.56$ signals the phase transition to the ferromagnetic state, while the shoulder at $T\sim 0.7$ is due to the development of short-range spin correlations. 
The square of magnetization $M^2=\braket{(\sum_{i} S^z_i/N)^2}$ is developed below the critical temperature, as plotted in Fig.~\ref{Fig:MCresult}(b). 
These bulk quantities are almost independent of $D$, as we consider a single impurity. 

In order to examine the local property around the impurity, we calculate the vorticity $l$ defined by 
\begin{align}
l=\frac{1}{2\pi} \sum_p (\phi_{p+1}-\phi_p), 
\end{align}
where $\phi_p$ is the $xy$-plane angle of spins measured from the $x$ axis, and $\phi_7 = \phi_1$. 
As shown in Fig.~\ref{Fig:MCresult}(c), the vorticity $\langle l \rangle$ becomes larger while increasing $D$ and decreasing $T$. 
It saturates to $\langle l \rangle=+1$ for $D=0.8$ at low $T$, which is consistent with the spin-wave result shown in Fig.~\ref{Fig:bindingenergy}(b). 
A Monte Carlo snapshot of the real-space spin configuration is shown in Fig.~\ref{Fig:MCresult}(d). 
Interestingly, the vorticity remains nonzero even for smaller $D\neq 0$ and above the critical temperature. 
This is presumably due to development of short-range correlations around the impurity.  
This suggests that the vortex state is expected even in the magnetic insulators with small spin-orbit coupling (small $D$), such as 3$d$ transition metal compounds, and even in the paramagnetic state.

\begin{table}[t!]
\caption{
Classification of twelve in-plane spin configurations around the nonmagnetic impurity with respect to the irreducible representation (irrep) of the magnetic point group $6m'm'$ and the corresponding cluster multipoles. 
Here, $S_p^x\propto(1,0,0)$ and $S_p^y\propto(0,1,0)$. 
See also Fig.~\ref{Fig:hole_ponti}(c). 
MM, MD, MQ, and MO represent magnetic monopole, dipole, quadrupole, and octupole, respectively. 
MTD, MTQ, and MTO represent magnetic toroidal dipole, quadrupole, and octupole, respectively. 
$l$ and $\gamma$ are the vorticity and helicity, respectively [Eq.~(\ref{eq:SxSy})].
}
\label{tab:irrep}
\centering
\begingroup
\renewcommand{\arraystretch}{1.2}
 \begin{tabular}{cclccc}
\hline \hline
$l$& $\gamma$ & spin patterns & irrep& multipole \\ \hline 
0& $0$, $\pi$ & $S^x_p$ & ${\rm E}_1$ & MD \\
& $\pm\pi/2$ & $S^y_p$ & &  \\
$\pm$3& $0$, $\pi$ & $ x_p (x_p^2-3y_p^2) S_p^x$ & ${\rm E_2}$ & MTO 
\\
& $\pm\pi/2$ & $x_p (x_p^2-3y_p^2)  S_p^y$ & & \\
$+$1& $0$, $\pi$ & $\bm{r}_p \cdot\bm{S}_p $ &  ${\rm A}_1$& MM \\
& $\pm\pi/2$&$(\bm{r}_p \times \bm{S}_p)^z $ & ${\rm A}_2$& MTD \\
$-$1& $0$, $\pi$ & $x_p S^x_p-y_p S^y_p $ & ${\rm E_2}$ & MQ \\
& $\pm\pi/2$ &$x_p S^y_p+y_p S^x_p$ &  & \\
$+$2& $0$, $\pi$ & $(x_p^2-y_p^2) S_p^x + 2x_py_p S_p^y$& ${\rm E}_1$ & MTQ \\
& $\pm\pi/2$ & $-2x_py_p S_p^x+(x_p^2-y_p^2) S_p^y$& &     \\
$-$2& $0$, $\pi$ & $(x_p^2-y_p^2) S_p^x - 2x_py_p S_p^y$ & ${\rm B}_2$ & MO \\
& $\pm\pi/2$ & $2x_py_p S_p^x+(x_p^2-y_p^2) S_p^y  $& ${\rm B}_1$ & MO \\
\hline 
\end{tabular}
\endgroup
\end{table}

\begin{figure}[t!]
\begin{center}
\includegraphics[width=1.0 \hsize]{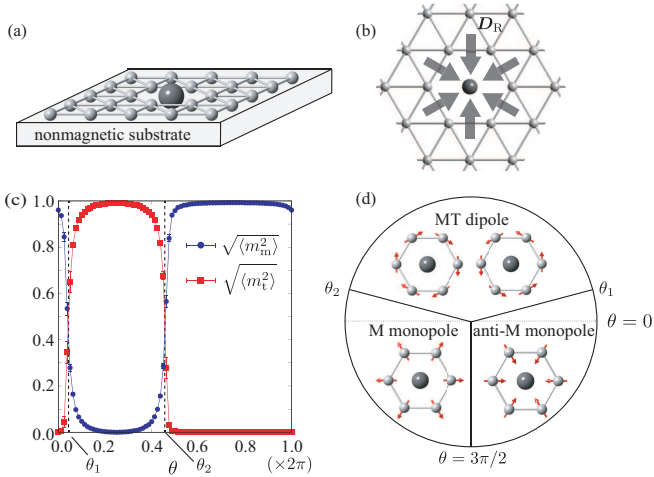} 
\caption{
\label{Fig:c6v}
(Color online) 
(a) Schematic picture of a monolayer ferromagnet on a nonmagnetic substrate. 
(b) Rashba-type DM interactions $D_{\rm R}$ induced by the mirror symmetry breaking with respect to the triangular plane. 
(c) $\theta$ dependences of the monopole and toroidal components, $m_{\rm m}$ and $m_{\rm t}$, respectively, at $T=0.01$, $J_3=-0.2$, $D=1$, $A=0.05$, and $R=0.1$. 
The crossovers between the M monopole and MT dipole states take place at $\theta_1 \sim 0.088 \pi$ and $\theta_2 = \pi - \theta_1$.
(d) Optimal spin structures classified by the active multipoles. 
}
\end{center}
\end{figure}

So far, we have considered a free-standing monolayer ferromagnet. 
Now, we discuss a potential realization of the magnetic vortex in a more realistic situation. 
Specifically, we consider the monolayer ferromagnet on a nonmagnetic substrate, as schematically shown in Fig.~\ref{Fig:c6v}(a). 
In this situation, reflecting the mirror symmetry breaking with respect to the plane, two types of additional interactions are induced: one is the Rashba-type antisymmetric interaction with in-plane DM vectors [Fig.~\ref{Fig:c6v}(b)] and the other is the anisotropic symmetric interaction~\cite{kaplan1983single}. 
Such exchange interactions can be taken into account by the Hamiltonian, 
\begin{align}
\label{eq:Ham_c6v}
\mathcal{H}_{\rm R}=&D_{\rm R}\sum_{p}  \bm{R}_p \cdot (\bm{S}_p \times \bm{S}_{p+1}) \nonumber \\
                           &+G_{\rm R} \sum_{p} \left[2(\bm{R}_p\cdot \bm{S}_{p})(\bm{R}_p\cdot \bm{S}_{p+1}) - \bm{S}_p \cdot \bm{S}_{p+1} \right], 
\end{align}
where $\bm{R}_p=\hat{\bm{z}}\times (\bm{r}_{p+1}-\bm{r}_{p})$ ($\hat{\bm{z}}$ is the unit vector in the $z$ direction).
$D_{\rm R}$ and $G_{\rm R}$ are the coupling constants for the in-plane Rashba-type antisymmetric and the anisotropic symmetric interactions, respectively, which are parameterized by $(D_{\rm R}, G_{\rm R})=R(\cos \theta, \sin \theta)$. 

As the term $\mathcal{H}_{\rm R}$ breaks the $U(1)$ symmetry in spin space around the impurity site, it lifts the degeneracy with respect to the helicity. 
The vortex states with particular helicity can be classified in terms of the cluster type multipoles. 
In the present case, arbitrary spin textures are characterized by the multipoles under the magnetic point group $6m'm'$~\cite{Suzuki_PhysRevB.95.094406,hayami2018microscopic,suzuki2018first,Suzuki_PhysRevB.99.174407}. 
For the six spins around the impurity, there are twelve independent $xy$ spin structures, as shown in Fig.~\ref{Fig:hole_ponti}(c). 
They are classified by the irreducible representations as ${\rm A}_1 \oplus {\rm A}_2 \oplus {\rm B}_1 \oplus {\rm B}_2 \oplus 2{\rm E}_1 \oplus 2{\rm E}_2$; each irreducible representation is associated with a particular cluster multipole~\cite{Hayami_PhysRevB.98.165110,Suzuki_PhysRevB.99.174407}. 
The multipole representations of the six-spin cluster are summarized in Table~\ref{tab:irrep}.

The vortex state with $l=+1$ obtained in the above calculations is characterized by the magnetic monopole (MM) when the helicity is $0$ or $\pi$ or by the magnetic toroidal dipole (MTD) when the helicity is $\pm \pi/2$. 
The MM and MTD are monitored by 
\begin{align}
m_{\rm m}=\frac{1}{6}\sum_{p}\bm{r}_p \cdot\bm{S}_p, 
\quad
m_{\rm t}=\frac{1}{6}\sum_{p}(\bm{r}_p \times \bm{S}_p)^z, 
\end{align}
respectively.
The states with the helicity 0 ($\pi$) and $\pi/2$ ($-\pi/2$) denote the (anti-)MM state with a positive (negative) $m_{\rm m}$ and (anti-)MTD state with a positive (negative) $m_{\rm t}$, respectively. 
Figure~\ref{Fig:c6v}(c) shows the Monte Carlo results for $\theta$ dependences of $\sqrt{\langle m^2_{\rm m} \rangle}$ and $\sqrt{\langle m^2_{\rm t} \rangle}$ at $T=0.01$, $J_3=-0.2$, $D=1$, $A=0.05$, and $R=0.1$. 
Three regions are distinguished by the dominant multipole components while changing $\theta$: 
the MTD state with the helicity $\pm \pi/2$ ($|m_{\rm t}|>0$) for $\theta_1<\theta<\theta_2$, the MM state with the helicity $0$ ($m_{\rm m}>0$) for $\theta_2<\theta<3\pi/2$, the anti-MM state with the helicity $\pi$ ($m_{\rm m}<0$) for $3\pi/2<\theta<0$ and $0<\theta<\theta_1$, as shown in Figs.~\ref{Fig:c6v}(c) and \ref{Fig:c6v}(d), where $\theta_1 \sim 0.088 \pi$, and $\theta_2 = \pi-\theta_1$. 
The results indicate that the large positive anisotropic symmetric exchange interaction $G_{\rm R}$ stabilizes the MTD, while the Rashba-type DM interaction $D_{\rm R}$ prefers MM or anti-MM. Furthermore, the negative (positive) $D_{\rm R}$ favors the (anti-)MM. We note that the results are anticipated from the symmetry of the interactions in Eq.~(\ref{eq:Ham_c6v}).
Thus, the observation of spin texture around the impurity will be useful for identifying the relevant microscopic exchange interactions. 

To summarize, we have elucidated that a nonmagnetic impurity doped into ferromagnets can nucleate a magnetic vortex for less frustration in exchange interactions and even in the absence of an external magnetic field. 
The mechanism lies in the emergence of asymmetric DM interactions around the impurity due to the breaking of inversion symmetry. 
By performing the spin-wave analysis and Monte Carlo simulations for the spin model on the triangular lattice, we found a stable bound state with the vorticity $+1$ in the wide parameter range. 
In the case of a monolayer ferromagnet on a nonmagnetic substrate, we found that the vortex state with particular helicity is stabilized, which is classified by the magnetic cluster multipoles. 
We showed that the magnetic toroidal dipole is induced for large positive anisotropic and small Rashba-type exchange couplings, while the magnetic monopole is in the rest parameter region. 
Our finding will provide a new way to nucleate, annihilate, and control a topological defect in magnets with the spin-orbit coupling, which can be applied to a variety of systems, such as monolayer metals on substrates and ferromagnetic/nonmagnetic heterostructures~\cite{elmers1995ferromagnetic,vaz2008magnetism}. 

In the presence of a finite density of impurities, the vortices begin to interact with each other.
An interesting issue is cooperative phenomena between such vortices.
A glassy state of magnetic vortices has been found in Ni$_x$Mn$_{1-x}$TiO$_3$ as a toroidal glass~\cite{Yamaguchi_PhysRevLett.108.057203,yamaguchi2013magnetoelectric}. 
Although the present results are not directly applied to this case, our mechanism related with symmetry lowering by randomness may play an important role in stabilizing such a state. 
Our results will serve as a good starting point for understanding of such randomness-induced cooperative phenomena in terms of multipoles.

\begin{acknowledgments}
S. H. thanks C. D. Batista, S.-Z. Lin, and Y. Kamiya for useful discussions. 
This research was supported by JSPS KAKENHI Grants Numbers JP15H05885, JP16H06590, JP18H04296 (J-Physics), and JP18K13488, and JST CREST (JPMJCR18T2). 
Parts of the numerical calculations were performed in the supercomputing systems in ISSP, the University of Tokyo. 
\end{acknowledgments}

\bibliographystyle{JPSJ}
\bibliography{ref}

\end{document}